\documentclass[
  twoside,
  twocolumn,
  aps,
  10pt,
  prb,
  floatfix,
  showpacs,
  citeautoscript,
]{revtex4-1}

\usepackage{amsmath}
\usepackage{amssymb}
\usepackage{graphicx}
\usepackage{tabularx}
\usepackage{dcolumn}
\usepackage[dvipsnames]{xcolor}
\usepackage{units}
\usepackage[acronym]{glossaries}

\usepackage[colorlinks=true]{hyperref}
\hypersetup{
    pdffitwindow=false,     
    pdfstartview={FitH},    
    pdfnewwindow=true,      
    colorlinks=true,
    linkcolor=NavyBlue,
    citecolor=NavyBlue,
    filecolor=NavyBlue,
    urlcolor=NavyBlue,
}

\setacronymstyle{long-short}

\newacronym{aic}{AIC}{alkaline information criteria}
\newacronym{ardr}{ARDR}{automatic relevance determination regression}
\newacronym{bcc}{BCC}{body-centered cubic}
\newacronym{bic}{BIC}{Bayesian information criterion}
\newacronym{cs}{CS}{compressive sensing}
\newacronym{cv}{CV}{cross-validation}
\newacronym{dft}{DFT}{density functional theory}
\newacronym{dof}{DOF}{degrees of freedom}
\newacronym{dos}{DOS}{density of states}
\newacronym{eam}{EAM}{embedded atom method}
\newacronym{fc}{FC}{force constant}
\newacronym{lasso}{LASSO}{least absolute shrinkage and selection operator}
\newacronym{adlasso}{ad-LASSO}{adaptive LASSO}
\newacronym{md}{MD}{molecular dynamics}
\newacronym{ols}{OLS}{ordinary least squares}
\newacronym{pes}{PES}{potential energy surface}
\newacronym{rfe}{RFE}{recursive feature elimination}
\newacronym{rmse}{RMSE}{root-mean-square error}

\newcommand{\hiphive}{\textsc{hiphive}}
\newcommand{\shengbte}{\textsc{shengBTE}}
\newcommand{\phonopy}{\textsc{phonopy}}
\newcommand{\phonothreepy}{\textsc{phono3py}}
\newcommand{\sklearn}{\textsc{scikit-learn}}

\newcommand{\alamode}{\textsc{ALAMODE}}
\newcommand{\ase}{\textsc{ASE}}

\newcommand{\bgg}{Ba$_8$Ga$_{16}$Ge$_{30}$}
\renewcommand{\vec}[1]{\ensuremath\mathbf{#1}}
\renewcommand{\epsilon}[0]{\varepsilon}

\begin{document}

\title{
    Efficient construction of linear models in materials modeling
    \texorpdfstring{\\}{}
    and applications to force constant expansions
}

\author{Erik Fransson}
\author{Fredrik Eriksson}
\author{Paul Erhart}
\email{erhart@chalmers.se}
\affiliation{
    Chalmers University of Technology,
    Department of Physics,
    Gothenburg, Sweden
}

\begin{abstract}
Linear models, such as force constant (FC) and cluster expansions, play a key role in physics and materials science.
While they can in principle be parametrized using regression and feature selection approaches, the convergence behavior of these techniques, in particular with respect to thermodynamic properties is not well understood.
Here, we therefore analyze the efficacy and efficiency of several state-of-the-art regression and feature selection methods, in particular in the context of FC extraction and the prediction of different thermodynamic properties.
Generic feature selection algorithms such as recursive feature elimination with ordinary least-squares (OLS), automatic relevance determination regression, and the adaptive least absolute shrinkage and selection operator can yield physically sound models for systems with a modest number of degrees of freedom.
For large unit cells with low symmetry and/or high-order expansions they come, however, with a non-negligible computational cost that can be more than two orders of magnitude higher than that of OLS.
In such cases, OLS with cutoff selection provides a viable route as demonstrated here for both second-order FCs in large low-symmetry unit cells and high-order FCs in low-symmetry systems.
While regression techniques are thus very powerful, they require well-tuned protocols.
Here, the present work establishes guidelines for the design of protocols that are readily usable, e.g., in high-throughput and materials discovery schemes.
Since the underlying algorithms are not specific to FC construction, the general conclusions drawn here also have a bearing on the construction of other linear models in physics and materials science.
\end{abstract}

\maketitle

\section{Introduction}

Linear models such as \gls{fc} and cluster expansions are widely used in materials science, physics, and chemistry to describe the thermodynamic behavior of real materials.
Their computational efficiency and mathematical simplicity are also appealing
for applications in high-throughput calculations and machine learning.
To this end, one requires methods for efficient and automatized model construction.
In this context, regression techniques are particularly appealing as they promise to require fewer computationally demanding reference calculations than approaches based on systematic enumeration of configurations \cite{EsfSto08, EsfLia19}.

Regression techniques in combination with regularization have received a lot of attention for model building, often under the title \gls{cs} \cite{CanWak08, NelHarZho13}.
The latter is in principle a task in sparse signal recovery that is usually approached by finding solutions to an underdetermined linear system.
The problem of solving the linear system is, however, completely independent of \gls{cs} and \gls{cs} itself is not a solver.

The usefulness of regression with regularization for the construction of physical models has been demonstrated, using \gls{lasso} \cite{TadTsu15} as well as the split-Bregman technique \cite{NelOzoRee13, ZhoNieXia14, AngLinErh16, ZhoNieXia19, ZhoSadAbe19}.
There is, however, a much larger pool of potentially useful regression techniques, including various other forms of regularization and feature selection.
These models involve one or several hyperparameters the choice of which often has a very direct impact on the results (as shown extensively below).
With applications in high-throughput computations but also more conventional situations in mind, it is therefore necessary to conduct a careful analysis of these techniques for the construction of physical models \cite{PlaNatUsa17}.

The vibrational \glspl{dof} of materials are crucial for numerous thermodynamic properties, including phase stability and thermal conduction \cite{Gri99, GriMagOzo12}.
To model these properties one requires an efficient representation of the \gls{pes}.
In crystals the vibrational atomic motion can be conveniently described in terms of phonons, quasi-particles that represent periodic and quantized excitations \cite{Wal98}.
Phonon theory is commonly formulated by starting from a Taylor expansion of the total energy, in which the expansion coefficients are referred to as \glspl{fc}.
Depending on material and property of interest the \gls{fc} expansion must be carried out to different orders.
Generally, it is preferable to keep the order as low as possible since the number of independent coefficients quickly increases with expansion order, decreasing symmetry, and number of sites in the unit cell \cite{EriFraErh19}.

For ideal materials with comparably small unit cells the \glspl{fc} up to third order can still be obtained by enumerating displacements and evaluating the derivatives numerically.
This direct enumeration scheme becomes, however, tedious or impractical for larger systems (e.g., point defects, interfaces or nanoparticles \cite{AgoAlb09, SopKotAlb11, KatCarDon17}) and/or materials that require expansions beyond third order (e.g., metastable phases of transition metals or oxides \cite{ThoVan13}).
Accordingly linear regression techniques have been applied including \gls{ols} \cite{EsfSto08, HelAbrSim11, HelSteAbr13, TadGohTsu14}, \gls{lasso} \cite{TadTsu15}, and split-Bregman \cite{ZhoNieXia14, ZhoNieXia19}.
As noted above, there are various other linear regression techniques and feature selection algorithms that could be suitable for \gls{fc} regression such as \gls{rfe}, \gls{ardr}, and \gls{adlasso}.
Further analysis of these techniques with regard to their efficiency, accuracy and reliability for constructing \gls{fc} expansions is therefore in order \cite{PlaNatUsa17}.

Here, we use the \hiphive{} package \cite{EriFraErh19, hiphive} since it is interfaced with machine learning libraries such as \sklearn{} that in turn provide efficient implementations of various optimization techniques.
In this paper we present a comparison of linear regression methods and the direct enumeration approach for the extraction of \glspl{fc} of different order, including second-order \glspl{fc} for large systems of low symmetry such as defects (\autoref{sect:tantalum-vacancy}), third-order \glspl{fc} for the prediction of the thermal conductivity (\autoref{sect:silicon-thermal-conductivity}), as well as higher-order \glspl{fc} for bulk (\autoref{sect:clathrate}) and surface (see Supplementary Information) systems.
This approach enables us to determine the applicability of regression methods in different regimes.
We also demonstrate the application of these \gls{fc} models for studying anharmonic effects, both in the framework of Boltzmann transport theory and \gls{md} simulations.
The following section provides a concise summary of the underlying theory, while sections thereafter present the different application examples named above.

\section{Force constant extraction}

The \gls{pes} can be expanded in a Taylor series in the atomic displacements $\vec{u}$ relative to a set of reference positions $\vec{r}_0$
\begin{align*}
    V &=
    V_0
    + \Phi_{i}^{\alpha} u_{i}^{\alpha}
    + \frac{1}{2}\Phi_{ij}^{\alpha\beta} u_{i}^{\alpha}u_{j}^{\beta}
    + \frac{1}{3!}\Phi_{ijk}^{\alpha\beta\gamma} u_{i}^{\alpha}u_{j}^{\beta}u_{k}^{\gamma}
    + \dots,
\end{align*}
where $\Phi$ are the \glspl{fc}, Latin indices enumerates the atoms, Greek indices enumerate the Cartesian coordinates, and the Einstein summation convention applies.

The number of \gls{fc} components scales as $\mathcal{O}(N^n)$, where $N$ is the number of atoms and $n$ is the expansion order.
There are, however, multiple constraints that reduce the number of free parameters, such as lattice symmetries and sum rules \cite{EriFraErh19}.
Yet in the case of large systems, low symmetry, and/or higher expansion orders the number of parameters is still very large.

\subsection{Direct approach}
\label{sect:direct-approach}

The conventional way of extracting \glspl{fc} relies on the systematic evaluation of numerical derivatives \cite{ParLiKaw97}.
For example, for the second-order terms
\begin{align*}
    \Phi_{ij}^{\alpha\beta}
    &=
    \frac{\partial^2 V}{\partial u_{i}^{\alpha} \partial u_{j}^{\beta}}
    \approx
    -\frac{F_i^\alpha}{\Delta u_{j}^{\beta}},
\end{align*}
where $F_i^\alpha$ denotes the force on atom $i$ along $\alpha$ and $\Delta u_{j}^{\beta}$ is a small displacement of atom $j$ along $\beta$, typically between 0.01 and \unit[0.05]{\AA}.
This \emph{direct approach} is implemented in several software packages, including \phonopy{} \cite{TogTan15} for second-order and \phonothreepy{} \cite{TogChaTan15}, \shengbte{} \cite{LiCarKat14}, \textsc{almabte} \cite{CarVerKat17}, and \textsc{AAFLOW} \cite{PlaNatUsa17} for third-order \glspl{fc} as well as \alamode{} \cite{TadGohTsu14} for an arbitrary expansion order.
This method has been used with great success for predicting vibrational properties of many common materials.
The number of reference calculations, however, quickly becomes a limiting factor for systems with many sites in the unit cell, in the case of low symmetry, and/or higher-order \glspl{fc}.
In fact, it is usually impractical to compute any term beyond third-order using the direct approach except for rather simple cases \cite{MadSan05}.

\subsection{Regression approach}
\label{sect:regression-approach}

\subsubsection{Linear form}

The information density, here taken as the number of force components that are sizable, in supercells with only one or two displaced atoms, such as the ones used in the direct approach, is relatively low.
Instead, one can consider general displacement patterns, involving many (or all) atoms in the supercell, and then employ regression techniques to reconstruct the underlying \glspl{fc}.
This approach has been shown to produce accurate higher order \glspl{fc} \cite{LinBroFra19, EsfCheSto11, ZhoNieXia14, ZhoNieXia19, ZhoSadAbe19, TadTsu18a, TadTsu18b, EsfSto08} but can also be used to construct effective \gls{fc} models \cite{HelAbrSim11, And12, HelSteAbr13, LinBroFra19}.

The force acting on atom $i$ along $\alpha$  can be written as
\begin{align*}
    F_i^\alpha
    &=
    -\Phi_{ij}^{\alpha\beta} u_{j}^{\beta}
    -\frac{1}{2}\Phi_{ijk}^{\alpha\beta\gamma}u_{j}^{\beta}u_{k}^{\gamma}
    -\dots,
\end{align*}
which can be cast in linear form \cite{EriFraErh19}
\begin{align*}
    F_i^\alpha = \mathbf{A}_i^\alpha \cdot \vec{x}.
\end{align*}
Here, $\vec{x}$ are the free parameters of the \gls{fc} model while the rows of the fit matrix $\mathbf{A}$ encode the displacements with symmetry transformations as well as constraints imposed by the sum rules.
The vector comprising all forces in a supercell $\vec{F}$ can thus be expressed as
\begin{align}
    \vec{F} &= \mathbf{A} \vec{x}.
    \label{eq:force_disp_linear_problem}
\end{align}
The construction of the fit matrix $\mathbf{A}$ is described in Ref.~\onlinecite{EriFraErh19} and can be trivially generalized to multiple reference structures.

\subsubsection{Truncating the expansion}
\label{sect:truncating-the-expansion}

The number of free parameters, i.e. the dimension of $\vec{x}$, can still be very large even for systems with high symmetry.
The \gls{fc} expansion is therefore often truncated.
Firstly, as with most Taylor expansions, only few orders are usually needed to obtain an accurate representation of the \gls{pes} in the range of relevant displacements.
Secondly, the atomic interactions often decay rather quickly with interatomic distance, meaning a cutoff can be imposed.
Thirdly, pair interactions are often stronger than three-body interactions, which in turn are often stronger than four-body interactions etc., i.e.
\begin{align*}
    \left \| \Phi_{iijj} \right \| > \left \| \Phi_{iijk} \right \| > \left \|\Phi_{ijkl} \right \|.
\end{align*}

\subsubsection{Linear regression techniques}

Equation \eqref{eq:force_disp_linear_problem} can be solved by minimizing the objective function $\left\Vert\mathbf{A}\vec{x}-\vec{F}_\text{target}\right\Vert_2$, where $\vec{F}_\text{target}$ denotes the reference forces.
In the overdetermined limit this can be achieved by \gls{ols} \cite{EsfSto08}, which can, however, lead to overfitting.\footnote{
    One should emphasize that the rows of the sensing matrix $\mathbf{A}$ in Eq.~\eqref{eq:force_disp_linear_problem} are always to some extent correlated due to atomic interactions as each row corresponds to the Cartesian force component of one atom in one structure.
    This should be kept in mind when using the terms overdetermined and underdetermined in the usual sense that is based on the relation between the number of rows and columns of the sensing matrix.
}

Overfitting can be overcome by regularization, i.e. inclusion of additional penalty terms in the objective function, usually related to the $\ell_1$ or $\ell_2$ norm of the solution vector.
For example in the case of elastic net regularization one has
\begin{align}
    \mathbf{x}_\text{opt}
    &=
    \underset{\mathbf{x}}{\text{argmin}}
    \left\{
        \left\Vert
            \mathbf{A} \mathbf{x} - \mathbf{F}_\text{target}
        \right\Vert^2_2
        + \alpha \left\Vert \mathbf{x} \right\Vert_1
        + \beta  \left\Vert \mathbf{x} \right\Vert^2_2
    \right\}.
    \label{eq:elastic-net}
\end{align}
With $\alpha=0$ this expression reduces to Ridge regression, while with $\beta=0$ one recovers the objective function for the \gls{lasso} method.
As discussed below, \gls{lasso} is known to over-select features.
To overcome this deficiency the \gls{adlasso} approach \cite{Zou06} has been proposed, in which the regularization term is modified to
\begin{align}
    \mathbf{x}_\text{opt} = \underset{\mathbf{x}}{\text{argmin}} \left \{ \left\Vert \mathbf{A} \mathbf{x}- \mathbf{F}_\text{target} \right\Vert^2_2 + \alpha \sum_i w_i  | x_i | \right \},
    \label{eq:adlasso}
\end{align}
where $w_i$ are individual weights for each parameter.
Here, we employ the iterative update described in Ref.~\onlinecite{Gramfort2012}.

To evaluate the performance of a model obtained by solving Eq.~\eqref{eq:force_disp_linear_problem} one can employ \gls{cv}.
To this end, the available reference data set is split into training and validation sets.
After the former has been used for fitting the parameter vector, one can evaluate the \gls{rmse}
\begin{align*}
    \text{RMSE}
    &=
    \sqrt{ \frac{1}{N}
    \sum_i \left(F_i^\text{model} - F_i^\text{target}\right)^2},
\end{align*}
where the summation extends over the $N$ force components in the validation set.
To reduce the statistical error, the \gls{rmse} is then averaged over several different splits of the reference data, yielding the \gls{cv} score.

Efficient and generally applicable implementations of these methods are available, e.g., via the Python machine learning library \sklearn{} \cite{PedVarGra11}.

\subsubsection{Feature selection}

In machine learning feature selection refers to the task of isolating the most important parameters (or features) during model construction.
Reducing the number of parameters yields a less complex model, which in turn often leads to less overfitting and improved transferability.
It can also reduce the computational cost of sampling the model.
Feature selection is especially interesting for \gls{fc} models, for which only some interaction terms may be of importance (\autoref{sect:truncating-the-expansion}).

Several feature selection methods are available for linear problems.
One can employ for example a simple pruning condition based on the magnitude of the parameters.
This is particularly effective in combination with regression techniques that include regularization, typically via the $\ell_1$ or $\ell_2$-norm of the parameter vector (see eq.~\eqref{eq:elastic-net}).
One can also employ matching pursuit algorithms such as orthogonal matching pursuit, which allows one to impose a constraint on the number of non-zero coefficients in the solution vector, or techniques such as \gls{rfe}.
In the latter case, a solution is determined using a fit method of choice, the weakest or least important parameters are removed, and the procedure is iteratively repeated until the target number of features is reached.
In some cases, the optimal sparsity of a model can be determined by combining the above techniques with Bayesian optimization \cite{NelOzoRee13}.

It must be noted that the different methods can differ considerably with respect computational effort as well as memory requirements.
\gls{ols} is the least demanding procedure in both regards.
It is difficult to provide general guidelines with respect to the demands of different methods, since the effort can differ dramatically with the choice of hyperparameters and the conditioning of the sensing matrix.
As a rough guideline, \gls{rfe} with \gls{ols} typically requires about 100 to 1,000 \gls{ols} fits depending on how accurately one wishes to perform the feature elimination (\autoref{fig:Ta_vacancy_fit_scaling}).
\gls{lasso} is comparable to \gls{rfe}-\gls{ols} with respect to computational effort.

\section{Second-order force constants: Large low-symmetry systems}
\label{sect:tantalum-vacancy}

The second-order \glspl{fc} of systems with many atoms and/or low symmetry can be tedious to obtain by the direct approach (\autoref{sect:direct-approach}).
This applies in particular to the \glspl{fc} of defect configurations, which are needed for example for computing the vibrational contribution to the free energy of defect formation \cite{AgoAlb09}, analyzing the impact of defects on the thermal conductivity \cite{KatCarDon17, DonCarKat18, KunOttCar19} or predicting the vibrational broadening of optical spectra \cite{AlkBucAws14, ShaHasBer20}.
In this section, we therefore analyze the extraction of second-order \glspl{fc} for the vacancy in \gls{bcc} Ta as a prototypical case, using both the direct approach (\autoref{sect:direct-approach}) as implemented in \phonopy{} \cite{TogTan15} and the regression approach  (\autoref{sect:regression-approach}) as implemented in \hiphive{} \cite{EriFraErh19}.

\subsection{Calculation of reference forces}

Calculations were carried out for supercells comprising $N\times N\times N$ conventional \gls{bcc} cells with $N\in[4, \ldots 10]$ that contain a single vacancy.
Reference forces were computed using the \gls{eam} potential model TA1 from Ref.~\onlinecite{RavGerGue13}.
Using an empirical potential rather than \gls{dft} calculations in this example allows us to compute reference second-order \glspl{fc} for very large configurations.
For the \phonopy{} calculations we used a displacement amplitude of \unit[0.01]{\AA} while for the \hiphive{} calculations, we generated 30 structures for each $N$ by drawing random displacements from a normal distribution with a standard deviation of \unit[0.01]{\AA}.

\subsection{Scaling of regression methods}

\begin{figure}
  \includegraphics{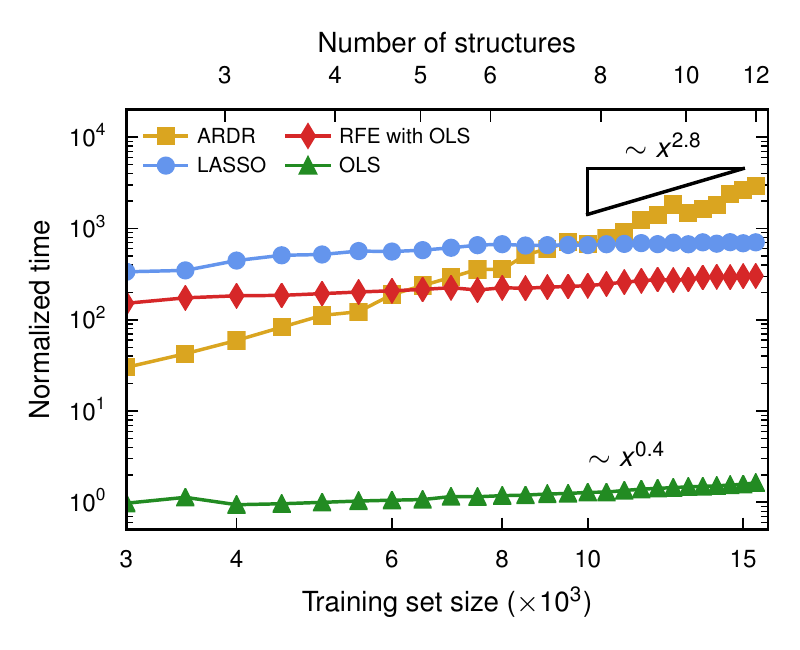}
  \caption{
      \textbf{Computational cost of different optimization algorithms.}
      Relative timings for the computational effort for carrying out a single optimization of second-order force constant models for a vacancy in Ta ($N=6$) with a cutoff of  \unit[6.0]{\AA} using different regression techniques.
      Time is normalized using the time for ordinary least squares (OLS) with a training set size of 5000 force components ($\sim$3\,seconds).
      Calculations were carried out on an Intel Xeon E5-2650 V3 processor with 10 cores (20 threads).
      ARDR: automatic relevance detection regression;
      LASSO: least absolute shrinkage and selection operator;
      OLS: ordinary least squares; 
      RFE: recursive feature elimination.
  }
  \label{fig:Ta_vacancy_fit_scaling}
\end{figure}

Several different methods were considered for constructing \gls{fc} models for the Ta vacancy models, including \gls{ols}, \gls{rfe}-\gls{ols}, \gls{lasso}, and \gls{ardr}.
\Gls{ols} is by far the computationally least expensive method and exhibits a favorable scaling with training set size (\autoref{fig:Ta_vacancy_fit_scaling}).
(We note that \sklearn{} employs the singular-value decomposition routine provided by \textsc{scipy} to solve the \gls{ols} problem, which in turn relies on \textsc{lapack}.)
\Gls{rfe}-\gls{ols} and \gls{lasso} exhibit a very similar scaling as \gls{ols} but are about 100 to 500 times more expensive.
This is unsurprising as these methods carry out multiple \gls{ols} optimizations as part of their algorithms.
Finally, \gls{ardr} exhibits an unfavorable scaling with training set size (including both computational effort and memory requirements), which prevents its effective application for large sensing matrices including the present case.

Taking into account \gls{cv}, the computational effort required for the largest supercells considered here becomes notable for \gls{lasso} and \gls{rfe}-\gls{ols}.
In the remainder of this section, we therefore primarily consider \gls{ols}, which will be demonstrated to work very well if combined with cutoff selection to avoid overfitting.

Furthermore, it is possible to tune parameters such as the rate at which the number of features is reduced in the \gls{rfe} algorithm in each step, which effectively reduces the pre-factor of \gls{rfe}.
The present comparison focuses, however, on the ability of these methods to recover physically correct solutions and their convergence behavior, whence aspects concerning their  computational performance are not further explored here.

\begin{figure}
  \includegraphics{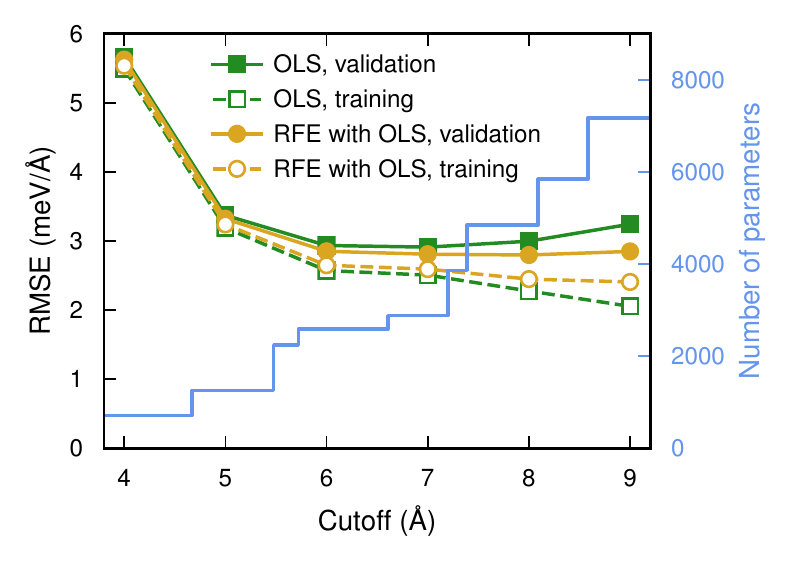}
  \caption{
      \textbf{Convergence with respect to parameter space for second-order Ta vacancy force constant models.}
      The plot shows the variation of the root-mean-square error (RMSE) over training and validation sets with pair cutoff and hence the number of degrees of freedom in the model.
      Calculations were carried out using 15 training structures based on a $6\times6\times6$ conventional supercell, corresponding to a total of 19,395 force components.
      OLS: ordinary least squares; 
      RFE: recursive feature elimination.
  }
  \label{fig:Ta_vacancy_cutoff_convergence}
\end{figure}

\subsection{Cutoff selection via cross-validation}

The number of \glspl{dof} in a \gls{fc} model grows rapidly as the cutoff increases.
At the same time, as discussed in \autoref{sect:truncating-the-expansion}, interaction strength and hence the magnitude of the \glspl{fc} decay with increasing interatomic distance.
At some point an increase in cutoff will therefore lead to negligible improvement in accuracy but merely an increase in model complexity.
Specifically in the absence of regularization terms in the objective function ($\alpha=\beta=0$ in Eq.~\eqref{eq:elastic-net}), one can therefore observe a deterioration of model quality with the inclusion of more terms in the expansion due to overfitting.
In this case one should therefore evaluate the performance of models with different cutoffs.

We employed \gls{cv} using the shuffle-and-split method with 5 splits and 15 training structures for a system size of $N=6$ and constructed a series of second-order \gls{fc} models with increasing cutoffs (\autoref{fig:Ta_vacancy_cutoff_convergence}).
While the \gls{rmse} over the training set continues to decrease with increasing parameter space, the \gls{cv}-\gls{rmse} has a minimum around 6 to \unit[7]{\AA}.
For standard \gls{ols} the validation score increases for larger cutoffs due to overfitting.
This behavior can be counteracted by using \gls{rfe}, which yields a slight improvement of the \gls{cv}-\gls{rmse}.
As discussed above \gls{rfe} is, however, computationally substantially more expensive whence \gls{ols} with a judicious choice of cutoffs is preferable.
All subsequent analysis was therefore carried out using \gls{ols} and a second-order cutoff of \unit[6]{\AA}.

\subsection{Convergence of thermodynamic properties}

\begin{figure}
  \includegraphics{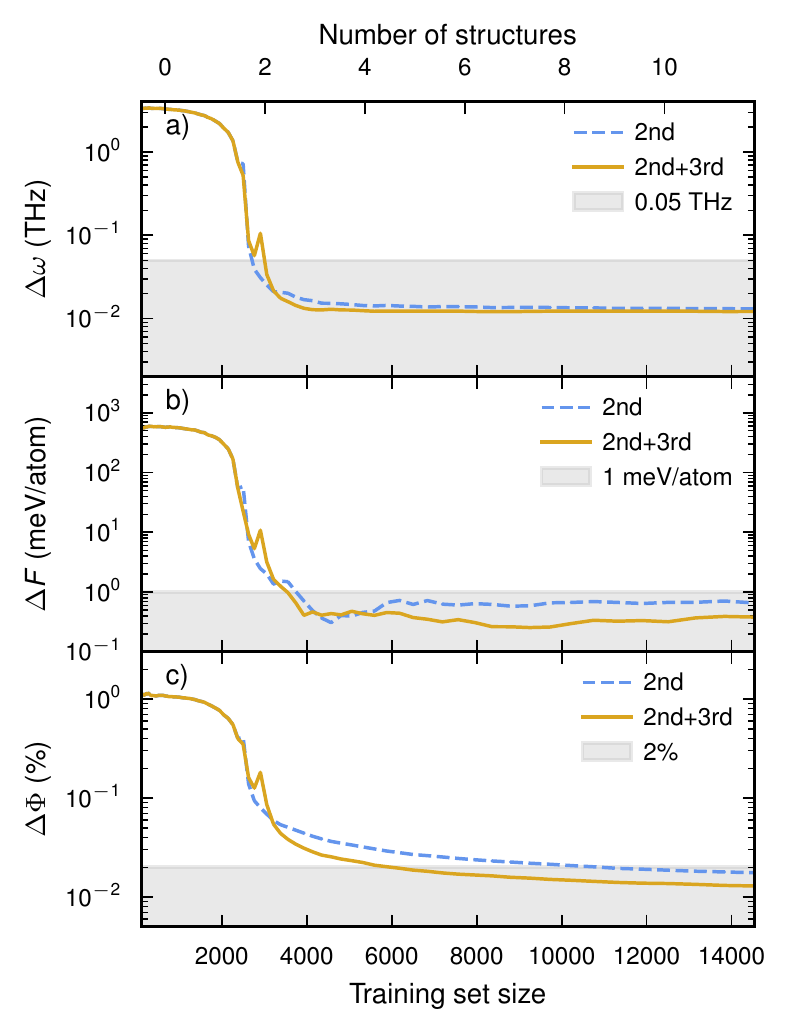}
  \caption{
      \textbf{Convergence with training set size for thermodynamic properties from Ta vacancy force constant models.}
      (a) Zone-center frequencies according to Eq.~\eqref{eq:error_omega}.
      (b) Free energy at 75\%\ of the calculated melting temperature (\unit[2250]{K}) according to Eq.~\eqref{eq:error_freen}.
      (c) Elements of the second-order \gls{fc} matrix according to Eq.~\eqref{eq:error_phi}.
      All calculations were carried out using a second-order cutoff radius of \unit[6]{\AA} (compare \autoref{fig:Ta_vacancy_cutoff_convergence}).
  }
  \label{fig:Ta_vacancy_property_convergence}
\end{figure}

In order to evaluate how accurately the regression approach reproduces the correct second-order \glspl{fc}, we considered three different measures.
First, we evaluated the absolute error of the zone-center ($\Gamma$) frequencies obtained by regression relative to the direct approach (\autoref{fig:Ta_vacancy_property_convergence}a),
\begin{align}
    \Delta \omega &= \sqrt{\frac{1}{3N}\sum_i ^{3N} (\omega_{\text{regression}} - \omega_{\mathrm{direct}})^2},
    \label{eq:error_omega}
\end{align}
where $\omega_i$ is the frequency of mode $i$.
Secondly, we considered the absolute difference in the harmonic free energy at \unit[2250]{K} corresponding to 75\%\ of the calculated melting temperature (\autoref{fig:Ta_vacancy_property_convergence}b),
\begin{align}
    \Delta F &= \left |F_\text{regression}^\text{vib} - F_\text{direct}^\text{vib}\right |.
    \label{eq:error_freen}
\end{align}
Here, the free energies were computed within the harmonic approximation using \phonopy{} \cite{TogTan15}.

Lastly, in order to obtain a computationally cheaper measure, we computed the relative error of the second-order \gls{fc} matrices (\autoref{fig:Ta_vacancy_property_convergence}c), defined as follows
\begin{align}
    \Delta \Phi &= \| \Phi_{\text{regression}} - \Phi_{\text{direct}}\| \big/ \|\Phi_{\text{regression}}\|
    \label{eq:error_phi}
\end{align}
where $\| \ldots \|$ denotes the Frobenius norm.

The frequencies and free energies exhibit very similar convergence behavior.
Both quantities reach convergence at about 3,000 force components, which is equivalent to two to three configurations and corresponds to the number of parameters in the model.
Comparison with the measure based on the \gls{fc} matrix itself, eq.~\eqref{eq:error_phi}, suggests that $\Delta\Phi\lesssim5\%$ is sufficient to achieve convergence of the frequency spectrum and the free energy.
Considering the convergence of $\Delta\Phi$ itself suggests a more conservative threshold of 2\%.

The comparison includes both models with only sec\-ond-order \glspl{fc} terms and models with additional very short-ranged third-order \gls{fc} terms using a cutoff of \unit[3.0]{\AA}.
The latter perform consistently better than the second-order-only models.
The inclusion of a few third-order terms thus stabilizes the extraction of the second-order \glspl{fc}, an observation that has also been made in other situations \cite{EriFraErh19, ZhoNieXia19}.
These terms enable one to account for anharmonicity in the vicinity of the reference positions that would otherwise be effectively included in the second-order \glspl{fc}.
This principle can also be applied to higher-order terms, where we have found that adding a few terms of the respective next-higher order yields more accurate \glspl{fc}.

\subsection{Scaling with system size}

\begin{figure}
  \includegraphics{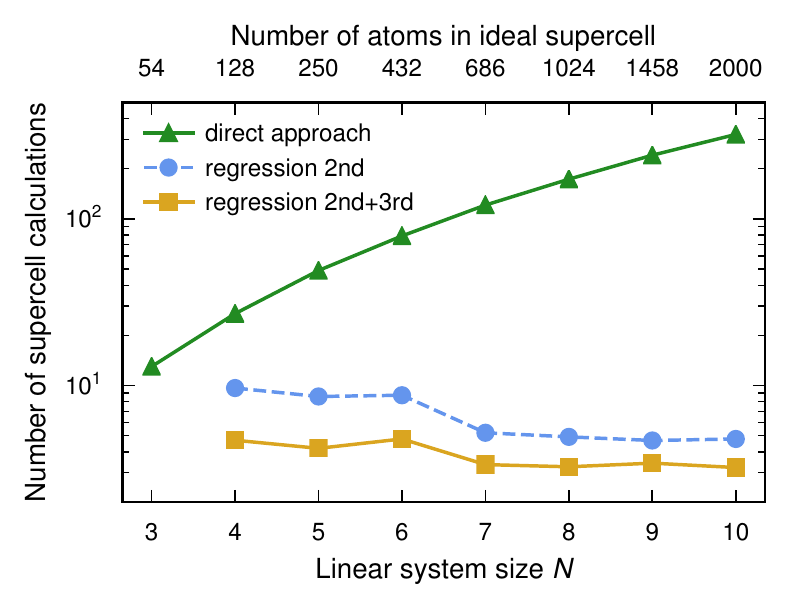}
  \caption{
      \textbf{Size scaling when extracting second-order Ta vacancy force constant models.}
      In the case of the regression approach, the force constant expansion include either second-order terms only or both second-order and a few (short-ranged) third-order terms.
    }
    \label{fig:Ta_vacancy_number_of_calculations}
\end{figure}

Following the analysis in the previous section, computing $\Delta \Phi$ as a function of the training set size allows one to determine the number of training structures needed for recovering the second-order \glspl{fc} at the accuracy level of the direct approach (\autoref{fig:Ta_vacancy_number_of_calculations}).
Using the more conservative threshold of $\Delta\Phi<2\%$ (\autoref{fig:Ta_vacancy_property_convergence}), we thereby determined the necessary number of supercell calculations as a function of system size (\autoref{fig:Ta_vacancy_number_of_calculations}).

While in the direct approach the number of necessary calculations increases steeply with system size, in the regression approach it is constant or decreases slightly with system size.
This is possible due to the introduction of a cutoff but more importantly since the regression takes advantage of the increase in information content with supercell size.
In this context, it is important that \emph{all} atoms are displaced by least a small amount.
By contrast, the information content in configurations employed in the direct approach decreases substantially with system size as only \emph{one} atom is displaced at a time.
One can anticipate this scaling effect to be even more pronounced for third or higher-order \glspl{fc} due to the exponential increase of the number of parameters with order \cite{EriFraErh19}.

\section{Third-order force constants: Thermal conductivity}
\label{sect:silicon-thermal-conductivity}

Calculating the thermal conductivity using the linearized Boltzmann transport equation requires knowledge of the second and third-order \glspl{fc} \cite{Zim60}, providing a sensitive test for the extraction of higher-order \glspl{fc}.
Here, we analyze different regression methods for obtaining \glspl{fc} and the resulting thermal conductivity in silicon.
Specifically, we consider \gls{ols}, which has been used for the same purpose in Ref.~\onlinecite{EsfSto08}, \gls{lasso}, \gls{adlasso}, \gls{rfe}-\gls{ols}, and \gls{ardr} as implemented in \sklearn{}.
This comparison enables us to demonstrate the importance of studying convergence with respect to the choice of cutoffs and number of training structures as well as the selection of a suitable fit method.

\subsection{Computational details}
\label{sect:silicon-computational-details}

Reference second and third-order \glspl{fc} were calculated for 250-atom supercells ($5\times5\times5$ primitive unit cells) via the direct approach using \phonopy{} \cite{TogTan15} and \phonothreepy{} \cite{TogChaTan15}, respectively.
No cutoff was imposed during the calculation of the third-order \glspl{fc}, which therefore required 801 individual force calculations.

For the regression approach, we generated a total of 20 reference structures based on 250-atom supercells ($5\times5\times5$ primitive unit cells) with displacements drawn from a normal distribution yielding an average displacement amplitude of \unit[0.03]{\AA}.
For the computation of \gls{cv} scores we used the same splits throughout to enable a one-to-one comparison of the regression methods.

Since \gls{ardr} exhibits a stronger scaling with system size than the other methods (\autoref{fig:Ta_vacancy_fit_scaling}), the pruning hyper-parameter $\lambda_t$ was not optimized but set to a constant value of $\lambda_t=10^4$.

For \gls{ardr} we first constructed a second-order-only model and then trained a second and third-order to the remaining forces.
This $\Delta$-approach led to a significant improvement in accuracy and numerical stability for \gls{ardr} but did not improve the performance of the other fit methods.

For \gls{adlasso} the hyperparameter $\alpha$ was optimized \emph{once} using \gls{cv} and five training structures, after which the same value was used for all training set sizes.
While one could thus possibly obtain slightly better results for larger training sets, the present choice allows a substantial reduction of the computational effort.

Reference forces were obtained from \gls{dft} calculations using the projector augmented wave method \cite{Blo94, KreJou99} as implemented in \textsc{vasp} \cite{KreFur96a, KreFur96b} and an exchange-correlation functional \cite{PerBurErn96} based on the generalized gradient approximation.
The Brillouin zone was sampled using only the $\Gamma$-point.
The plane-wave energy cutoff was set to \unit[245]{eV}, an additional support grid for fast-Fourier transformations was used during the force calculation, and the projection operators were evaluated in reciprocal space.

\subsection{Optimization with cutoff selection}

The number of \glspl{dof} associated with the second-order \glspl{fc} is very small and thus we used the maximum cutoff range of \unit[9.65]{\AA} that the $5\times5\times5$ supercells employed here can support.
The nearest neighbor fourth-order interaction, corresponding to a fourth-order cutoff of \unit[2.5]{\AA}, was included in order to improve the accuracy of the second and third-order \glspl{fc} (as demonstrated in \autoref{sect:tantalum-vacancy}).
The third-order cutoff was then treated as a tunable parameter.

\begin{figure}
    \includegraphics{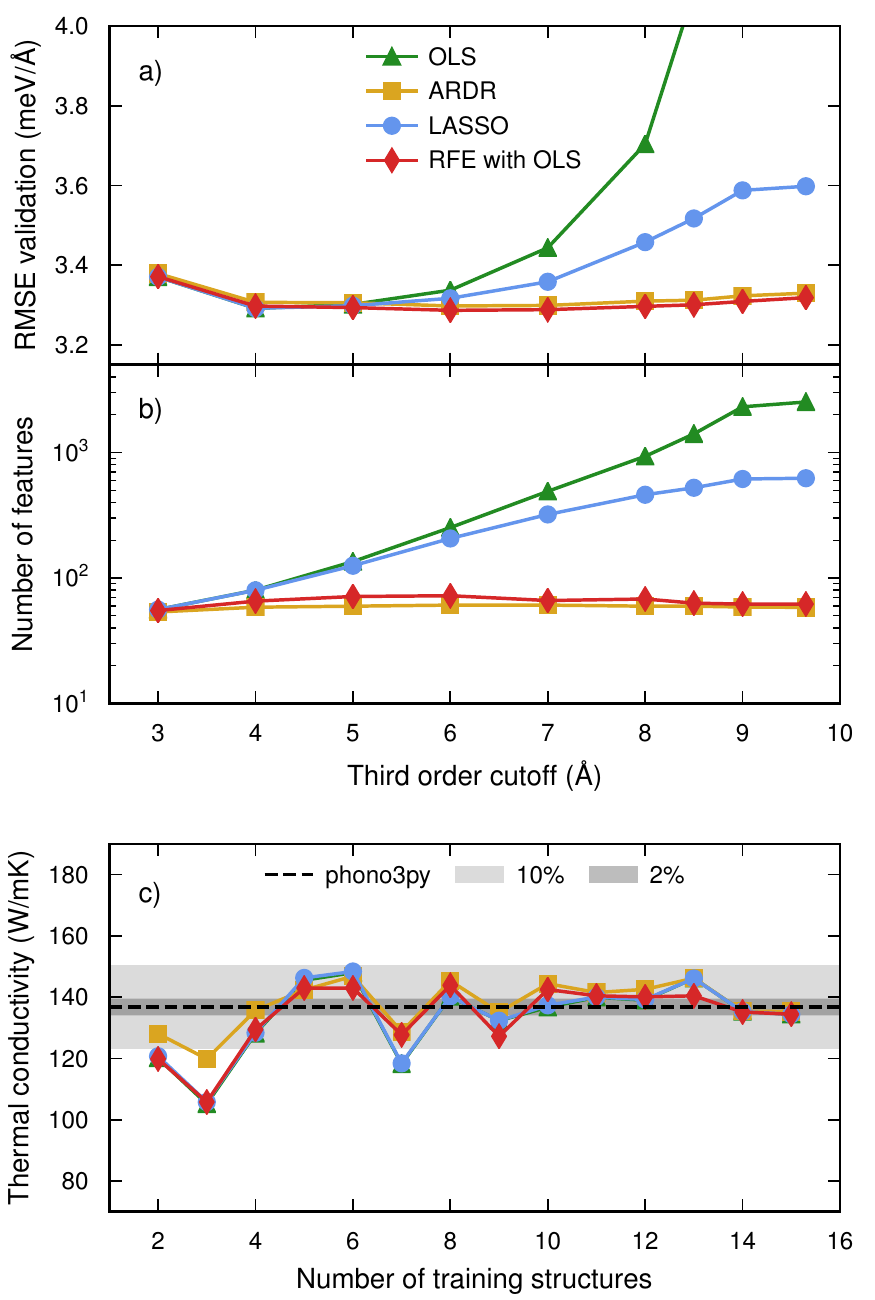}
    \caption{
        \textbf{Comparison of third-order force constant models for Si constructed using a small parameter space.}
        Convergence of (a) cross-validated root mean square error (RMSE) and (b) the number of features (non-zero parameters) with respect to third order cutoff using five training structures.
        Cutoffs of 9.65 and \unit[2.5]{\AA} were employed for second and fourth-order clusters, respectively.
        In ordinary least squares (OLS) no features are selected and hence in this case the number of features is identical to the total number of parameters.
        (c) Thermal conductivity at \unit[300]{K} as a function of number of training structures using a third-order cutoff of \unit[4.0]{\AA}.
        ARDR: automatic relevance detection regression;
        LASSO: least absolute shrinkage and selection operator;
        RFE: recursive feature elimination.
    }
    \label{fig:Si_summary1}
\end{figure}

When using five training structures the accuracy of the model is already converged for a third-order cutoff of \unit[4.0]{\AA} (\autoref{fig:Si_summary1}a) regardless of fit method, which yields a total of 80 \glspl{dof}.
\gls{rfe}-\gls{ols} and \gls{ardr} have, however, the distinct advantage of selecting fewer parameters and thus avoid overfitting for large cutoffs (\autoref{fig:Si_summary1}b).

Using a third-order cutoff of \unit[4.0]{\AA}, the thermal conductivity converges to within 2\%\ of the \phonothreepy{} values using as few as 14 structures (\autoref{fig:Si_summary1}).
For comparison, the \phonothreepy{} calculation requires 801 structures if no cutoff is imposed and 57 structures when including a pair cutoff of \unit[4.0]{\AA} (two neighbor shells).
In the latter case, one must, however, take into account that convergence testing would require including at least one more shell, which increases the number of calculations to  95 (three neighbor shells).

For example, in the case of high-throughput studies one is often content with a less accurate estimate of the thermal conductivity.
In this context, it is noteworthy that when using \gls{ardr} or \gls{rfe}-\gls{ols} one converges to within 10\%\ of the reference value already with four structures, while being able to test for convergence.

\subsection{Optimization with generic feature selection}

\begin{figure}
    \includegraphics{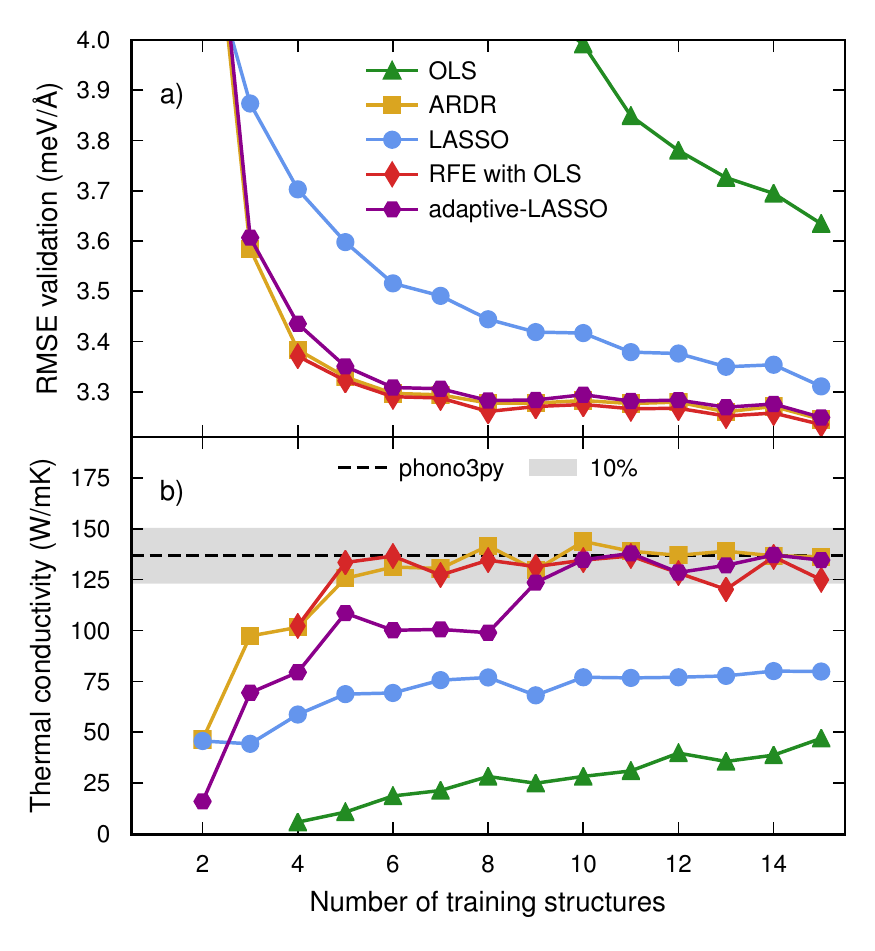}
    \caption{
        \textbf{Convergence of third-order force constant models for Si with training set size when using a large parameter space.}
        Convergence of (a) cross-validated root mean square error (RMSE) and (b) the thermal conductivity at \unit[300]{K} for generic feature selection methods as well as ordinary least squares (OLS) with respect to the number of training structures.
        Cutoffs of 9.65, 9.65 and \unit[2.5]{\AA} were employed for second, third and fourth-order clusters, respectively.
        ARDR: automatic relevance detection regression;
        LASSO: least absolute shrinkage and selection operator;
        RFE: recursive feature elimination.
    }
    \label{fig:Si_summary2}
\end{figure}

\begin{figure}
    \includegraphics{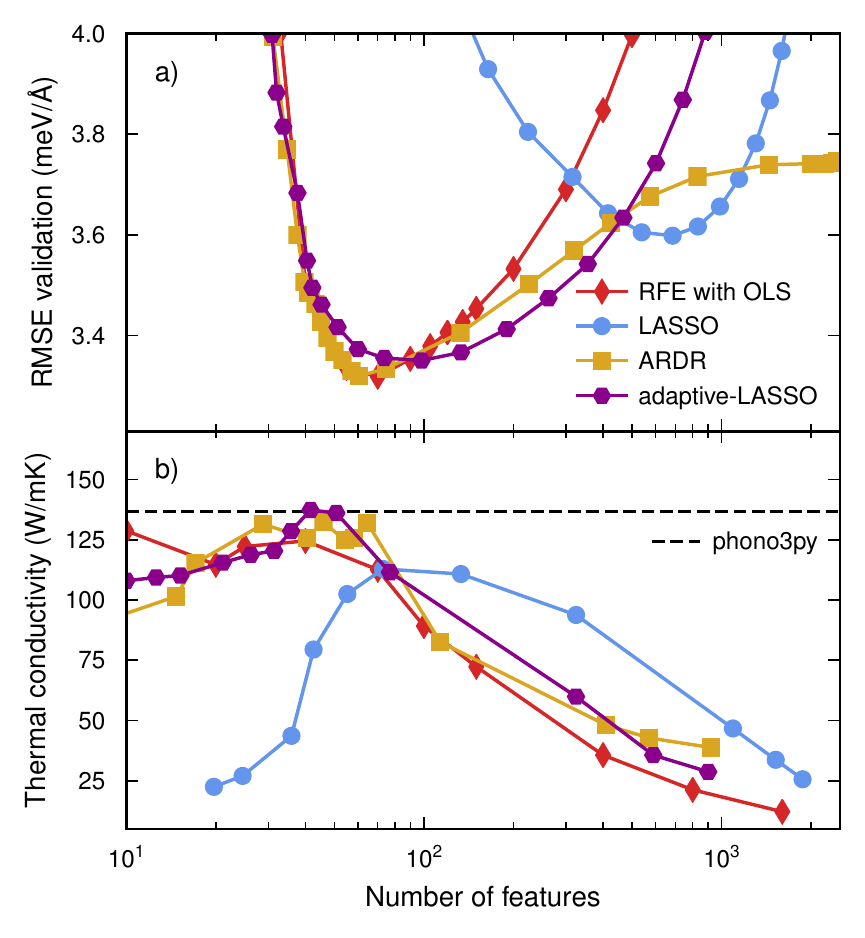}
    \caption{
        \textbf{Sparseness and accuracy of models obtained using different optimization methods.}
        Variation of (a) the cross-validated root mean square error (RMSE) and (b) the thermal conductivity at \unit[300]{K} with the number of features for Si third-order force constant models, constructed using five training structures and cutoffs of 9.65, 9.65 and \unit[2.5]{\AA} for second, third and fourth-order clusters, respectively.
        The number of features in each model was tuned via the hyperparameters of the respective optimization algorithms.
        The reference value for the thermal conductivity is shown by the dashed line in b and was computed using \textsc{phono3py}.
        ARDR: automatic relevance detection regression;
        LASSO: least absolute shrinkage and selection operator;
        RFE: recursive feature elimination.
    }
    \label{fig:Si_summary2b}
\end{figure}

Cutoff selection can be thought as a feature selection approach, in which the cutoffs are pruning hyperparameters.
In the present case, in which we focus on the third-order \glspl{fc} we effectively obtain only one such hyperparameter, namely the third-order cutoff.
The number of cutoff parameters increases with the expansion order and in more complex materials can require fine tuning e.g., between different atomic species or crystallographic sites \cite{LinBroFra19}.
In such situations it can be advantageous to employ generic feature selection algorithms.
In this section, we consider the suitability of \gls{lasso}, \gls{adlasso}, \gls{rfe}-\gls{ols} as well as \gls{ardr} for this purpose.
For comparison, we also include the performance of \gls{ols} is included.
Cutoffs of 9.65, 9.65, and \unit[2.5]{\AA} for second, third and fourth orders respectively were used throughout, which yields 2525 \glspl{dof}.

In the case of \gls{rfe}-\gls{ols} and \gls{ardr}, the thermal conductivity converges to within 10\%\ of the reference value using about five structures (\autoref{fig:Si_summary2}b; see Supplementary Figure~2 for temperature dependence).
Moreover, with \gls{ardr} one achieves convergence within 2\%\ with about 12 structures, which is even better than in case of cutoff selection.
In the case of \gls{lasso} and \gls{ols}, the convergence rate is considerably lower and neither method achieves convergence with respect to the reference value with the number of structures considered here.

We note that based on the convergence of the \gls{cv}-\gls{rmse} scores (\autoref{fig:Si_summary2}a) and the comparison with the cutoff selection study (\autoref{fig:Si_summary1}a), one could expect \gls{lasso} to yield a reasonably converged thermal conductivity when using about 15 structures.
This is not the case, demonstrating that \gls{cv}-\gls{rmse} scores \emph{alone} are insufficient for assessing the quality of a model.
Information criteria such as \gls{aic} and \gls{bic} can be used to evaluate models \cite{Aka74, Sch78, AhoDerPet14}, taking into account the predictive power but also penalizing the number of parameters of the model.
These types of measure may serve as a useful compliment to the \gls{cv} score when evaluating \gls{fc} models.
They were, for example, used recently to evaluate alloy cluster expansion models \cite{ZhaLiuBi20}.

\begin{figure}
    \includegraphics{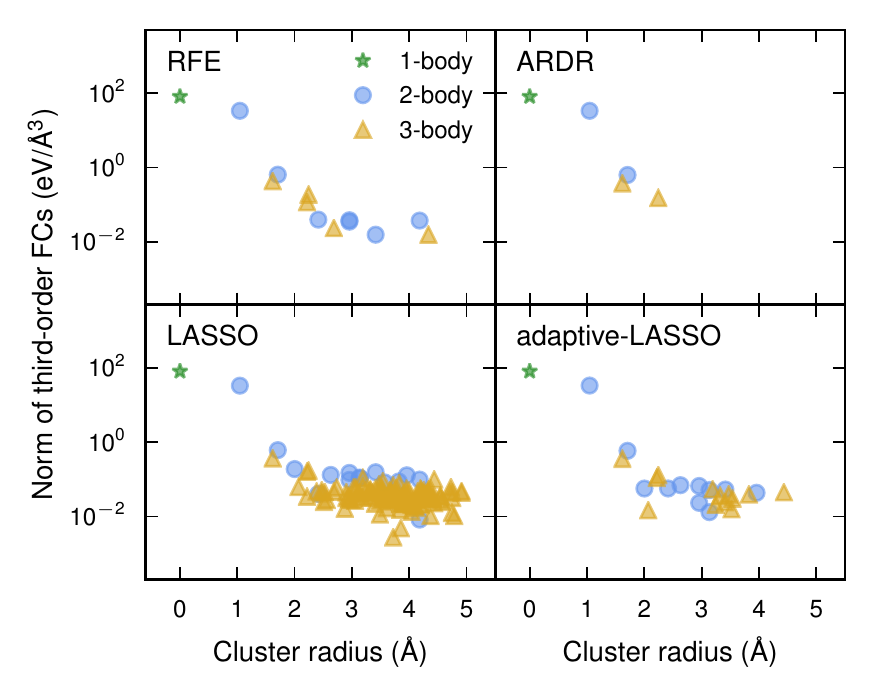}
    \caption{
        \textbf{Ability of different optimization methods to recover the correct third-order force constants.}
        Norm of third-order force constants in Si obtained using generic feature selection algorithms and ten training structures.
        Cutoffs of 9.65, 9.65 and \unit[2.5]{\AA} were employed for second, third and fourth-order clusters, respectively.
        The cluster radius is defined as $r_c = \sum_{i \in \text{cluster}}  \left \| \vec{r}_i - \vec{r}_{gc}  \right \| \big/ N$ where $\vec{r}_{gc}$ is the geometrical center of the cluster (compare Supplementary Figure~2).
        ARDR: automatic relevance detection regression;
        LASSO: least absolute shrinkage and selection operator;
        RFE: recursive feature elimination.
    }
    \label{fig:Si_fcs}
\end{figure}

To explore the differences in performance between \gls{lasso}, \gls{rfe}-\gls{ols}, and \gls{ardr} further we explicitly computed the \gls{cv}-\gls{rmse} as a function of their respective pruning hyperparameter using five training structures, which allows us to obtain the variation of the \gls{cv} score with the number of features (\autoref{fig:Si_summary2b}a).
While the methods achieve comparable \gls{rmse} scores, the optimal \gls{lasso} solution contains a much larger number of features.
The tendency of (standard) \gls{lasso} to over-select is known \cite{Zou06} and we have observed this behavior also in other applications such as alloy cluster expansions \cite{AngMunRah19}.
A physical understanding of the shortcoming of \gls{lasso} is obtained by inspecting the \glspl{fc} directly (\autoref{fig:Si_fcs}).
The second-order \glspl{fc} are very similar for all methods (not shown) but notable differences are observed in the third-order \glspl{fc}.
While \gls{rfe}-\gls{ols} and \gls{ardr} produce a small number of short-ranged interaction terms, \gls{lasso} yields a large number of spurious third-order \glspl{fc} terms (see Supplementary Figure~2).

The over-selection in \gls{lasso} can be overcome by using, e.g., \gls{adlasso}, see Eq.~\eqref{eq:adlasso} \cite{Zou06}.
This yields a learning curve comparable to \gls{ardr} (\autoref{fig:Si_summary2}a) and a much smaller number of features compared to \gls{lasso} (\autoref{fig:Si_summary2b}a).
Thereby, one obtains a small set of third-order terms (\autoref{fig:Si_fcs}) that properly reproduce the physical properties of the system (Fig.~S2).

It is striking that all techniques except for \gls{lasso} can achieve convergence of the thermal conductivity to within 5\%\ with only five training structures (\autoref{fig:Si_summary2b}b).
It is also noteworthy that the \gls{cv} score alone is an unreliable predictor for model quality.
This is especially apparent in the case of \gls{lasso}, for which the model with the  smallest \gls{cv} score leads to an underestimation of the thermal conductivity by 50\%.
Using the \gls{cv} score alone for model selection can hence be very misleading.
For this purpose, one could therefore also consider model evaluation metrics such as \gls{aic} and \gls{bic} -- an aspect that deserves further study.

Finally, we note that overestimating the true number of features leads to very large errors in the predicted thermal conductivity for all techniques.
Underestimation on the other hand, i.e. overly sparse models, give much smaller errors, which indicates that over-regularization is the preferable mode of error.

We emphasize that the data sets used here are publicly available \cite{gitlab_repo} and can serve as a test bed for a systematic comparison with respect to other fit algorithms and feature selection methods.

\section{Fourth-order force constants: Strong anharmonicity}
\label{sect:clathrate}

\begin{figure*}[bt]
  \includegraphics[scale=0.95]{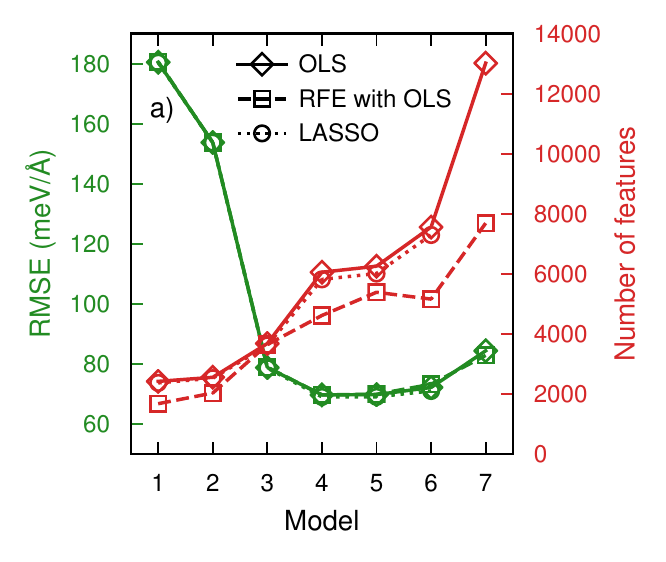}
  \includegraphics[scale=0.95]{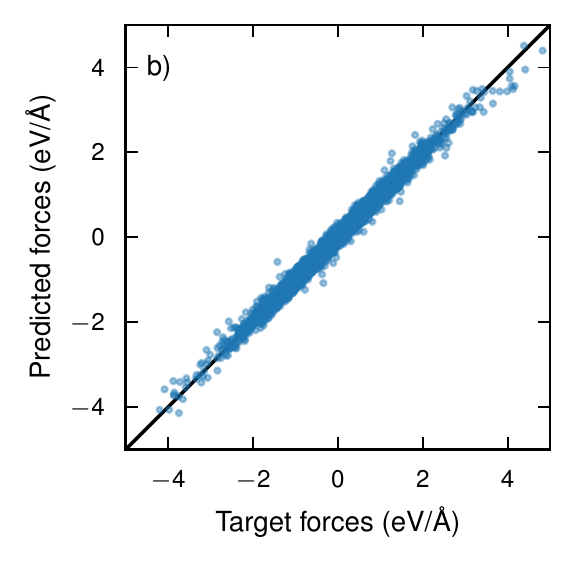}
  \includegraphics[scale=0.95]{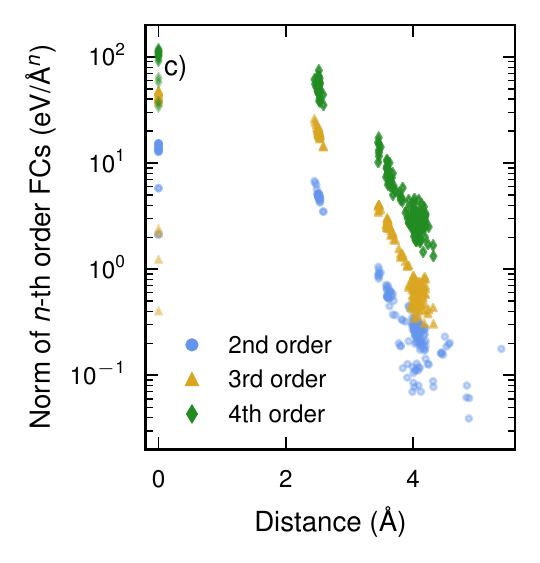}
  \caption{
    \textbf{Fourth-order \gls{fc} models for \bgg{}.}
    (a) Comparison of cross-validated root mean square error (RMSE) and number of features for the models defined in \autoref{tab:clathrate_models}.
    (b) Predicted vs target forces and (c) norm of the force constants (FCs) for Model 4.
    LASSO: least absolute shrinkage and selection operator;
    OLS: ordinary least squares;
    RFE: recursive feature elimination.
  }
  \label{fig:clathrate_models}
\end{figure*}

In this section, we are concerned with the inorganic clathrate \bgg{}, in which the motion of Ba atoms is strongly anharmonic \cite{MadSan05}.
This manifests itself in a strong temperature dependence of vibrational modes associated with Ba \cite{ChrAbrChr08} and moreover has implications for the thermal conductivity \cite{ChrAbrChr08, TadTsu18b, LinBroFra19}.
While perturbation theory formally provides an expression for the temperature induced phonon frequency shifts caused by the third-order \gls{fc} terms \cite{Wal98}, one commonly carries the expansion at least to the next higher \emph{even} order, when analyzing frequency shifts \cite{ErrCalMau14, TadTsu18b, LinBroFra19}.
Since \bgg{} has a large unit cells, it thus serves as an example for a system, in which both higher-order \glspl{fc} are required and the number of \glspl{dof} is very large.

Clathrates are inclusion compounds with a defined lattice structure that can trap atomic or small molecular species.
\bgg{} belongs to the class of type-I clathrates with spacegroup Pm$\bar{3}$n \cite{SheKov11}.
In this case, the host lattice is made up of Ga and Ge atoms, which occupy Wyckoff sites $6c$, $16i$, and $24k$, whereas Ba atoms reside inside the cages occupying Wyckoff sites $2a$ and $6d$.
Due to the size mismatch between guest species and cages, which is particular large for the $6d$ sites, the Ba atoms experience a very wide and flat \gls{pes} with pronounced anharmonicity.
In earlier work, we analyzed the ordering of the host species and extracted the ordered ground state structure of \bgg{} \cite{AngLinErh16, AngErh17}, which serves as a prototype structure for the analysis of the \glspl{fc}.

\subsection{Calculation of reference forces}

Reference forces were obtained for the 54-atom cells described below from \gls{dft} calculations using the computational approach described in \autoref{sect:silicon-computational-details}.
The exchange-correlation potential was represented using the vdW-DF-cx method, which combines semi-local exchange with non-local correlation \cite{DioRydSch04, KliBowMic11, BerHyl14}, as it has been previously shown to provide a good description of the vibrational modes of this system \cite{LinBroFra19}.
The Brillouin zone was sampled using a $\Gamma$-centered $3\times3\times3$ $\mathbf{k}$-point mesh and the plane-wave energy cutoff was set to \unit[243]{eV}.

\begin{table}[b]
    \centering
    \caption{
        Fourth-order \gls{fc} models for \bgg{} obtained by \gls{ols} using different combinations of cutoffs and expansion orders.
    }
    \label{tab:clathrate_models}
    \begin{tabular}{l *{5}d c *{3}d}
    \hline\hline
    \textbf{Model}
    & \multicolumn{5}{c}{\textbf{Two-body cutoffs}}
    && \multicolumn{3}{c}{\textbf{Three-body cutoffs}} \\
    \cline{2-6} \cline{8-10}
    & \multicolumn{1}{c}{2$^\text{nd}$}
    & \multicolumn{1}{c}{3$^\text{rd}$}
    & \multicolumn{1}{c}{4$^\text{th}$}
    & \multicolumn{1}{c}{5$^\text{th}$}
    & \multicolumn{1}{c}{6$^\text{th}$}
    &
    & \multicolumn{1}{c}{3$^\text{rd}$}
    & \multicolumn{1}{c}{4$^\text{th}$ order}
    \\
    \hline
    1     & 5.4 & 3.0  & 3.0  &     &             &&           \\
    2     & 5.4 & 3.5  & 3.5  &     &             &&           \\
    3     & 5.4 & 4.0  & 4.0  &     &             &&           \\
    4     & 5.4 & 4.35 & 4.35 &     &             &&           \\
    5     & 5.4 & 4.7  & 4.7  &     &             &&           \\
    6     & 5.4 & 4.35 & 4.35 & 3.0 & 3.0         &&           \\
    7     & 5.4 & 4.35 & 4.35 &     &             && 4.0 & 4.0 \\
    \hline\hline
\end{tabular}
\end{table}

\subsection{Model construction}
\label{sect:clathrate-model-construction}

First, we generated 50 structures based on the primitive 54-atom unit cell with an average displacement amplitude of \unit[0.28]{\AA} using the Monte Carlo rattle approach described in Ref.~\onlinecite{EriFraErh19}.
These structures were used to train an initial fourth-order model, which was subsequently sampled by \gls{md} simulations at 300 and \unit[650]{K} for \unit[10]{ps}.
We extracted 50 structures from each runs and generated reference forces via \gls{dft} calculations.
Thereby we obtained a total of 150 reference structures and thus a sensing matrix with 24,300 rows.
Due to the large number of \glspl{dof} in this structure (\autoref{tab:clathrate_models}) the computational effort that has to be expended for training models becomes very significant.
For illustration, a \gls{ols} fit with 10-fold \gls{cv} of the largest model considered below requires approximately one hour on an Intel Xeon E5-2650 V3 CPU.
Given the scaling analysis above (\autoref{fig:Ta_vacancy_fit_scaling}) this translates to several days and more when using \gls{rfe}-\gls{ols} or \gls{lasso}.
Since the computational effort associated with \gls{adlasso} is yet another order of magnitude larger than for \gls{lasso}, we restricted ourselves to \gls{ols}, \gls{rfe}-\gls{ols}, and (standard) \gls{lasso}, the latter of which has been previously used for a very similar structure \cite{TadTsu18b}.
We constructed several different models and evaluated their performance by \gls{cv} in order to identify a suitable combination of expansion order and cutoff parameters (\autoref{tab:clathrate_models}).

\begin{figure}
  \includegraphics{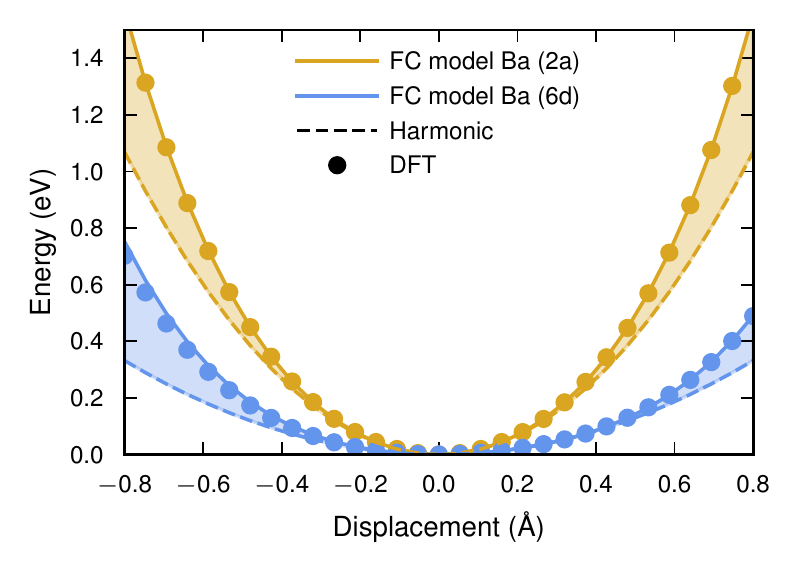}
  \caption{
    \textbf{Potential energy landscape for Ba atoms located at 2a and 6d Wyckoff sites along the $\left<111\right>$ direction.}
    The potential energy surfaces along $\left<100\right>$ and $\left<110\right>$ are shown in Supplementary Figure~5.
  }
  \label{fig:clathrate_pes}
\end{figure}

The three optimization methods considered here yield virtually identical \gls{cv}-\gls{rmse} scores (\autoref{fig:clathrate_models}a), which is sensible as even for the largest cutoff parameters the number of \glspl{dof} is still significantly smaller than the number of rows in the sensing matrix (\autoref{tab:clathrate_models}).
\Gls{rfe}-\gls{ols} and to a very slight extent also \gls{lasso} yield a smaller number of features.
In practice, one has to balance this advantage (which can also imply shorter run times for force evaluations of the final model) against the additional effort required for construction of the model.
Based on the good scores that can already be obtained with \gls{ols} and the fact that these cases are commonly easy to move into the overdetermined region, even for spaces as large as in the present case, one can argue that \gls{ols} is often the more natural choice.
In the following, we therefore restrict our analysis to the \gls{ols} fits.

The smallest \gls{cv}-\gls{rmse} score is obtained for model 4 (\autoref{fig:clathrate_models}a), which yields a value of \unit[68]{meV/\AA} to be compared with maximum force components of about \unit[4000]{meV/\AA} (\autoref{fig:clathrate_models}b).
The fourth-order cutoff for this model is \unit[4.35]{\AA}, which is enough to include all Ba-cage interactions indicating that the anharmonicity of all of these interactions is important.
The \gls{pes} for Ba atoms along different directions calculated is in excellent agreement with \gls{dft} calculations (\autoref{fig:clathrate_pes}).
It is apparent that Ba atoms in 6d sites behave more anharmonic than those in 2a sites, as the cages surrounding the former are larger.

To analyze the behavior of model 4 further, we generated an ensemble of \gls{fc} models that are trained in identical fashion but are based on different training sets.
The latter were constructed by selection with replacement (bagging) from the available data set such that the number of reference forces in the training sets equals the total size of the data set.
The \glspl{rmse} over the training set obtained for these \gls{fc} models are similar to model 4.
This ensemble of models enabled us to estimate the sensitivity for predictions for physical properties such as phonon frequencies (\autoref{fig:clathrate_dos}) and thermal conductivity (\autoref{fig:clathrate_kappa}).

\subsection{Thermal conductivity}

The low thermal conductivity of inorganic clathrates is often attributed to the rattling motion of the guest species \cite{MadSan05, ChrAbrChr08}.
These modes exhibit a notable temperature dependence \cite{ChrAbrChr08} that needs to be accounted for in order to predict the thermal conductivity accurately \cite{TadTsu18b, LinBroFra19}.
The common approach to Boltzmann transport theory, however, only considers terms up to third-order and neglects the temperature dependence of the phonon frequencies.
The fourth-order model allows us to investigate the temperature dependence of the frequency spectrum and include this effect when calculating the thermal conductivity via the temperature dependent force constants approach \cite{And12, HelAbr13, LinBroFra19}.
Yet instead of training effective third-order models from ab-initio \gls{md} simulations, we train them against snapshots and forces generated from \gls{md} simulations using the full fourth order model.
The latter simulations were carried out using a supercell of $2\times2\times2$ primitive unit cells (432 atoms).
The systems was first equilibrated for \unit[10]{ps} using a Langevin thermostat as implemented in \ase{} \cite{LarMorBlo17}.
Subsequently, the simulations were continued in the micro-canonical ensemble for another \unit[5]{ps}.
From the latter part, 100 snapshots were selected to train effective second and third-order \glspl{fc}.

\begin{figure}
  \includegraphics{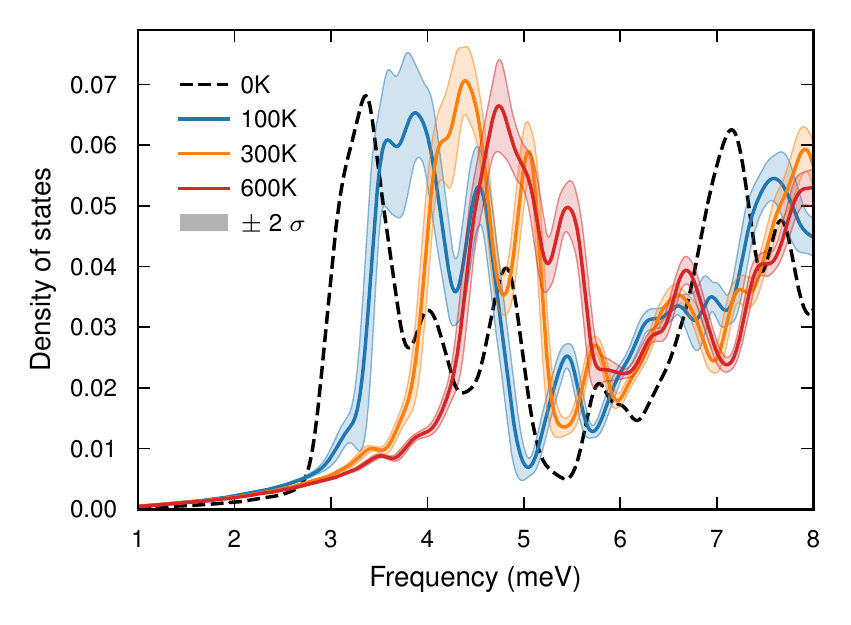}
  \caption{
    \textbf{Temperature dependent phonon density of states for \bgg{} from effective second-order force constants.}
    Shaded regions indicate $\pm 2\sigma$, for each respective temperature, obtained from the sensitivity analysis described in the text.
  }
  \label{fig:clathrate_dos}
\end{figure}

The \gls{dos} obtained from the effective second-order \glspl{fc} reveals a significant temperature dependence of the low-frequency Ba modes around \unit[4]{meV} (\autoref{fig:clathrate_dos}), in line with experimental work \cite{ChrAbrChr08} and our previous study, which was based on ab-initio \gls{md} simulations \cite{LinBroFra19}.
The sensitivity analysis (see \autoref{sect:clathrate-model-construction}) shows that the dependence of the \gls{dos} on model uncertainty is considerably weaker than the temperature dependence.
We also note that the \gls{dos} has a slightly weaker temperature dependency if the harmonic force constants are trained without including the third-order \glspl{fc}.
This is in qualitative agreement with the negative frequency shift due to cubic force constants observed in Ref.~\onlinecite{TadTsu18b}.

\begin{figure}
  \includegraphics{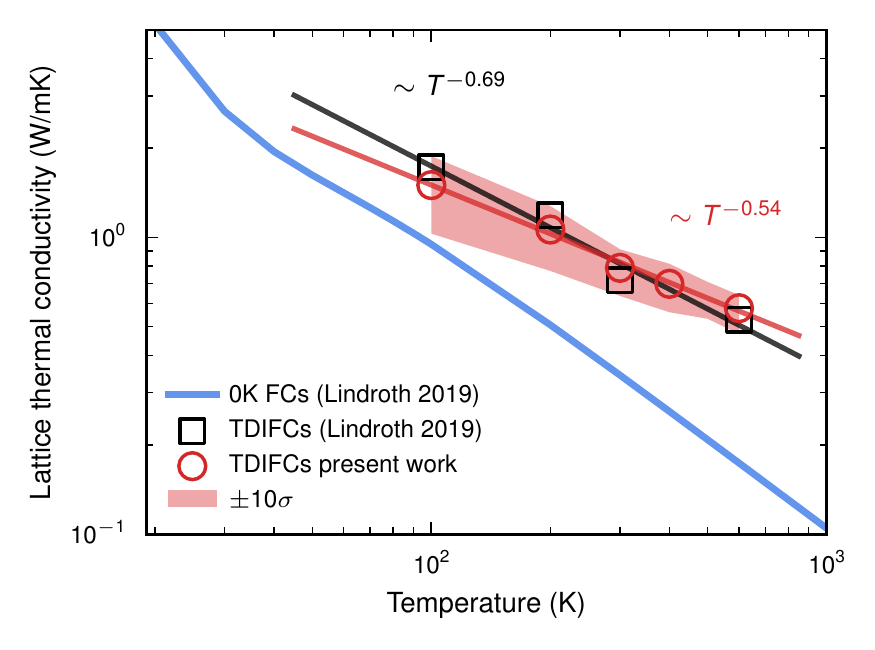}
  \caption{
    \textbf{Thermal conductivity for \bgg{} computed from the linearized Boltzmann transport equation.}
    Data shown by solid blue lines and open black square taken from Ref.~\onlinecite{LinBroFra19}.
    Shaded regions indicate $\pm 10\sigma$, for each respective temperature, obtained from the sensitivity analysis described in the text.
    TDIFCs: temperature-dependent interatomic force constants.
  }
  \label{fig:clathrate_kappa}
\end{figure}

The effective \glspl{fc} were furthermore used to extract the thermal conductivity at the respective temperature they were trained using the linearized Boltzmann transport equation via \shengbte{} \cite{LiCarKat14}.
The resulting thermal conductivity is in very good agreement with our previous calculations (\autoref{fig:clathrate_kappa}), which were based on full ab-initio \gls{md} simulations and which in turn agrees well with experimental results \cite{LinBroFra19}.
We note that as shown in Refs.~\onlinecite{TadTsu18b} and \onlinecite{LinBroFra19}, the thermal conductivity is strongly underestimated when using the temperature-independent (zero-K) second and third-order \glspl{fc} as that approach fails to account for the strong renormalization with temperature of the lowest-frequency heat carrying modes.
As in the case of the \gls{dos}, the model sensitivity analysis reveals a very weak variation in thermal conductivity due to model uncertainty.

The very good agreement with the results obtained using temperature-dependent interatomic force constants trained using ab-initio \gls{md} simulations demonstrates that our fourth-order model provides an accurate and transferable description of the \gls{pes} in the temperature range of interest.
The effective models in the present work, however, required only a fraction of the computer time that was needed for the ab-initio \gls{md} simulations in Ref.~\onlinecite{LinBroFra19}.

The present results are also in semi-quantitative agreement with the results from Ref.~\onlinecite{TadTsu18b}, which were obtained using a Green's function approach for treating phonon renormalization.
The latter calculations were carried using a different structural model that does not correspond to a thermally accessible state, whence one cannot expect quantitative agreement.
There are further methodological differences between the Green's function approach in Ref.~\onlinecite{TadTsu18b} and the present work.
The latter takes into account quantum statistic but only handles renormalization of the second-order \glspl{fc}, whereas the present approach is classical but effectively accounts for higher-order terms via temperature-dependent third-order \glspl{fc} \cite{LinBroFra19}.

\section{Discussion}

Regression techniques can reduce the computational effort associated with \gls{fc} extraction compared to enumeration techniques by one or more orders of magnitude, in particular in the case of higher-order expansions and/or large systems with low symmetry.
In this study, we have therefore assessed both efficiency and efficacy of several regression methods in different application scenarios.
Specifically, we considered second-order \glspl{fc} and derived properties for large systems (\autoref{sect:tantalum-vacancy}), third-order \glspl{fc} and thermal conductivity (\autoref{sect:silicon-thermal-conductivity}) as well as fourth and higher-order \glspl{fc} and their sampling for strongly anharmonic systems (\autoref{sect:clathrate} and Supplementary Information).

While the discussion below is explicitly based on the results presented in the preceding sections, they are further supported by our experience with various other materials including, e.g., metals, oxides, carbides, and chalcogenides of varying dimensionality including two and three-dimensional systems, interfaces, and defects.
Hence, we consider our conclusions to be applicable to other materials and application areas to a reasonable extent.

\subsection{Second-order \texorpdfstring{\glspl{fc}}{FCs} in large systems}

For second-order \glspl{fc} in large systems the regression approach can reduce the computational effort by more than one order of magnitude compared to the direct approach (\autoref{fig:Ta_vacancy_number_of_calculations}).
\Gls{ols} with cutoff selection yields prediction errors that are on par with more advanced regression techniques such as \gls{rfe}-\gls{ols}, \gls{lasso}, or \gls{ardr}.
The latter are, however, at least one to two orders of magnitude more demanding in terms of computer time, which can become a concern for very large systems.

For \gls{ols} to work properly the linear system to be solved must be overdetermined.
The configurations used for regression can be obtained by rattling the atomic positions.
As a result, the information density, i.e. the number of force components that are sizable, is high, which is usually not the case for enumerated structures such as the ones used in the direct approach.
As a result, a much smaller number of configurations is required in order to obtain a well conditioned sensing matrix.

We have furthermore found that inclusion of a few higher-order \gls{fc} terms (here third-order \glspl{fc} with a short cutoff) accelerates convergence of the lower-order \glspl{fc} of interest with respect to training set size.
Here, the third-order terms allow extraction of the ``true'' second-order expansion terms, which otherwise would have to effectively account for anharmonicity in the \gls{pes}.

\subsection{Third-order \texorpdfstring{\glspl{fc}}{FCs} and thermal conductivity}

The regression approach also drastically reduces the number of reference calculations needed to recover the parameters of third-order \gls{fc} expansions.
Here, care must be taken to verify not only the convergence of the \gls{cv}-\gls{rmse} scores with respect to the references forces but to consider the actual property of interest, in the present case the thermal conductivity.

As shown previously \cite{EsfSto08}, \gls{ols} with cutoff selection provides a viable route to obtaining well-converged thermal conductivity values at a fraction of the computational cost of the direct approach.
Several generic feature selection methods provide, however, viable alternatives that require adjusting a smaller number of hyperparameters and are hence more easily extensible to complex systems and higher-order expansions.

Here, \gls{rfe}-\gls{ols}, \gls{ardr}, and \gls{adlasso} have been found to work very reliably and efficiently, yielding both fast convergence and a small number of features (sparse solutions).
Standard \gls{lasso} produces denser solutions, which leads to less predictive models and has a detrimental impact when predicting the thermal conductivity.

\subsection{Higher-order \texorpdfstring{\gls{fc}}{FC} models and anharmonic \texorpdfstring{\glspl{pes}}{PESs}}

We also considered the construction of \gls{fc} expansions beyond third-order, which is usually impractical with enumeration approaches.
Specifically, we constructed fourth-order \gls{fc} models with up to more than 13,000 parameters for the inorganic clathrate \bgg{} using \gls{ols} with cutoff selection.
Using the final model effective \glspl{fc} were generated and used to compute the temperature dependent \gls{dos} and thermal conductivity, yielding results in agreement with experiment as well as previous computational work.

These results demonstrate that the ``standard'' and computationally relatively cheap \gls{ols} method can be used to capture strongly anharmonic effects across a wide temperature range.
It can easily be extend to higher order anharmonicity and properties other than the thermal conductivity, as shown by an eighth-order model for a Ni surface, which allows one to model the temperature dependence of the surface layer spacing (see Supplementary Information).

To put the present results in context, we note that the benefits of regression with regularization, typically discussed under the heading of \gls{cs}, have been particularly emphasized for systems with many \glspl{dof}, which includes both the clathrate system (over 13,000 parameters) and the Ni surface (2,000 parameters) considered here.
While for other linear models, specifically cluster expansions, it can be difficult to generate that many reference data points, for \gls{fc} expansions generating reference data is relatively easy and often simpler than tuning of model parameters and hyperparameters.
For example, for the clathrate system, one structure provides 162 reference force components and hence already with less than a hundred such structures one obtains an overdetermined system.

One could argue that regularization would allow one to reduce the number of reference data points considerably, making techniques such as \gls{adlasso}, \gls{lasso} or even \gls{ardr} competitive.
In this regard, we point out that the \gls{rfe}-\gls{ols} fits for the clathrate system show that a sensible model has about 4000 to 5000 parameters.
Even with regularization, one should thus require approximately as many reference data points, which is often not a substantial enough reduction to warrant the substantially larger computational effort associated with regularization techniques.

\section{Conclusions and outlook}

Generic feature selection algorithms, in particular \gls{ardr}, \gls{rfe}, and \gls{adlasso} can yield physically-sound \gls{fc} expansions at a fraction of the cost of enumeration approaches.
This approach can be very powerful as demonstrated here for extracting third-order \glspl{fc} and thermal conductivity.
The application of (standard) \gls{lasso} approach is, however, not indicated due to its tendency to over-select, which leads to very slow convergence with training set size.
For large unit cells with low symmetry and/or high-order expansions these techniques come, however, with a non-negligible cost that can be more than two orders of magnitude higher than that of \gls{ols}.
The cost is still much smaller than those of \gls{dft} calculations but since the underlying problem is not as amenable to parallelization it can still become a factor to consider in practice.
In such cases \gls{ols} with cutoff selection provides a viable route, with trivial parallelization over multiple cutoff parameter sets.
The viability of the latter approach has been demonstrated here for both second-order \glspl{fc} in large low-symmetry unit cells and high-order \glspl{fc} in low-symmetry systems.

Regression techniques are in principle very attractive for high-throughput schemes, since they require much less computational effort than enumeration approaches.
For the regression approach to be viable one must, however, not only consider the computational effort but also the amount of human intervention required.
In the future, it is therefore desirable to set up protocols that automatically construct and validate \gls{fc} models.

Finally, we have made the data related to the analysis of higher-order \gls{fc} expansions publicly available \cite{gitlab_repo} to provide a standardized benchmark set for future work.

\section*{Data and code availability}

The data and code related to the analysis of third-order models in Si and fourth-order models in \bgg{} is publicly available in the form of a \textsc{gitlab} repository at \url{https://gitlab.com/materials-modeling/hiphive-examples}.

\section*{Acknowledgments}

This work was funded by the Knut and Alice Wallen\-berg Foundation (2014.0226), the Swedish Research Council (2015-04153, 2018-06482), and the Swedish Foundation for Strategic Research (RMA15-0052).
Com\-puter time allocations by Swedish National Infrastructure for Computing at C3SE (Gothenburg), NSC (Link\"oping), and PDC (Stockholm) are gratefully acknowledged.


%

\end{document}